\documentclass[prd,aps,twocolumn,superscriptaddress,longbibliography]{revtex4-2}
\usepackage{amssymb}
\usepackage{amsmath}
\usepackage{graphicx} 
\usepackage{xcolor}
\usepackage{float}
\usepackage{hyperref}
\hypersetup{colorlinks=true, citecolor=blue, urlcolor=blue, linkcolor=blue,urlcolor=cyan}

\begin{document}

\title{Chaos-Integrability Transition in the BPS Subspace of the $\mathcal{N}=2$ SYK Model}

\author{Leon Miyahara}\email{leon@eken.phys.nagoya-u.ac.jp}
\author{Shono Shibuya}\email{shibuya.shono.n8@s.mail.nagoya-u.ac.jp}
\affiliation{Department of Physics, Nagoya University, Nagoya, Aichi 464-8602, Japan}

\begin{abstract}
We study a chaos-integrability transition purely within a BPS subspace, the subspace of exactly degenerate ground states annihilated by the supercharges, in a supersymmetric model that interpolates between the chaotic $\mathcal{N}=2$ SYK model and an integrable $\mathcal{N}=2$ ``commuting" SYK model. Since all BPS states are degenerate under the Hamiltonian, conventional Hamiltonian level statistics cannot diagnose chaos within the BPS subspace. Using the framework of BPS chaos, we instead analyze the spectrum of a simple operator projected onto the BPS subspace. 
We numerically find that its spectral statistics exhibit random-matrix behavior near the SYK regime and undergo a smooth crossover to Poisson statistics near the integrable regime. 
Our results provide a direct example in which the BPS data alone diagnose the chaos-integrability crossover.

\end{abstract}

\maketitle

\noindent \emph{1.~\textbf{Introduction}.—}
Quantum chaos plays a central role in understanding the quantum nature of black holes \cite{Sekino:2008he,Maldacena:2015waa}. 
The Sachdev-Ye-Kitaev (SYK) model \cite{Sachdev_1993,KitaevTalks,Maldacena:2016hyu} provides a tractable setting for studying this connection, as it exhibits quantum chaos, while its low-energy dynamics is closely related to nearly-AdS$_2$ gravity, known as Jackiw-Teitelboim (JT) gravity \cite{Jackiw:1984je,Teitelboim:1983ux}, through the Schwarzian effective theory \cite{Jensen:2016pah,Maldacena:2016upp}.
Various deformations of SYK models have been discussed to investigate chaos-integrability transition \cite{Garcia-Garcia:2017bkg,Nosaka:2018iat,Garcia-Garcia:2020cdo,Tezuka:2022mrr}, offering a useful laboratory to explore the relation between chaos and black hole dynamics.

Supersymmetric black holes provide a particularly valuable setting for microscopic studies of black hole physics; for instance, their protected entropies can often be counted from the dual CFT to match with an area of the horizon. However, supersymmetric black holes consist of Bogomol'nyi-Prasad-Sommerfield (BPS) states which are degenerate ground states, so the conventional diagnostics of quantum chaos based on spectral correlations of the Hamiltonian cannot be directly applied, in contrast to the belief that black holes should exhibit chaotic property.

A recent proposal, known as \emph{BPS chaos}, addresses this issue by considering the spectrum of a ``simple" operator $O$ projected onto the BPS subspace \cite{Lin:2022rzw,Lin:2022zxd,Chen:2024oqv},
\begin{equation}\label{eq:OBPS}
    O_{\rm BPS}=P_{\rm BPS} O P_{\rm BPS}\,.
\end{equation}
Random-matrix statistics in the spectrum of $O_{\rm BPS}$ are then interpreted as a signature of quantum chaos within the BPS subspace. 
Relatedly, the notion of \emph{fortuity} classifies BPS states according to whether they can be continued to larger $N$ \cite{Chang:2024zqi}. 
The associated \emph{chaos-invasion} picture suggests that the spectral properties of finite-energy states may be inherited by fortuitous BPS states \cite{Chen:2024oqv}.

These developments motivate a natural question: 
Can a chaos-integrability transition be detected solely in a BPS subspace?

In this work, we answer this question affirmatively in a deformed $\mathcal N=2$ supersymmetric SYK model. 
We consider a supercharge that interpolates between the usual chaotic $\mathcal N=2$ SYK supercharge \cite{Fu:2016vas} and an integrable ``commuting" supercharge. 
After confirming the chaos-integrability transition in the conventional setting of a non-BPS sector, we focus instead on a BPS subspace and study the spectrum of a projected operator. 
This provides a direct example of a chaos-integrability crossover diagnosed purely within a BPS subspace. Our results also suggest that BPS-sector diagnostics may be useful for probing the breakdown of black-hole-like chaotic descriptions in supersymmetric holographic systems.

\vspace{4pt}
\noindent \emph{2.~ $\mathbf{\mathcal{N}=2}$ \textbf{SYK model and its deformation}.—}
We consider the $\mathcal{N}=2$ supersymmetric SYK model with $N$ complex fermions $\psi_i$, whose supercharge is given by
\begin{equation}\label{eq:SYKQ}
    Q_{\rm SYK} = \sum_{i<j<k}^{N}C_{ijk}\psi_i\psi_j\psi_k\,,
\end{equation}
where the coupling constants $C_{ijk}$ are taken to be independent Gaussian random complex variables with zero mean and non-zero variance
\begin{equation}
    \overline{C_{ijk}C_{ijk}^*}=\frac{2}{N^2}\,.
\end{equation}
Then the Hamiltonian $H$ is given by an anticommutator
\begin{equation}\label{eq:H_SYK}
    H_{\rm SYK}=\{Q_{\rm SYK},Q_{\rm SYK}^{\dagger}\}\,.
\end{equation}
The R-charge associated with this Hamiltonian is the fermion number
\begin{equation}\label{eq:f}
    f=\sum_{i=1}^{N}\psi_i^{\dagger}\psi_i\,,
\end{equation}
and the Hilbert space is decomposed into sectors fixed by fermion number, denoted by $\mathcal{H}_f$.

For generic states $\Psi$, \eqref{eq:H_SYK} leads to the following inequality.
\begin{equation} \label{eq:Hbound}
    \langle\Psi|H_{\rm SYK}|\Psi\rangle=||Q_{\rm SYK}\Psi||^2+||Q_{\rm SYK}^\dagger\Psi||^2\geq0\,,
\end{equation}
which means that the eigenvalues of $H_{\rm SYK}$ are all non-negative, and this bound is called the \emph{BPS bound}. The states that satisfy this bound are zero-energy ground states and we call them \emph{BPS states} \footnote{Usually, they are called supersymmetric states because they preserve all the supercharges, i.e., $Q\Psi=Q^{\dagger}\Psi=0$, but we call them BPS states in a sense that they satisfy the BPS bound. Generally, when there are multiple supercharges, not all of them can be preserved at the same time. In fact, in a suitable basis, supercharges $\tilde Q_i$ should follow the relation $\{\tilde Q_i,\tilde Q_i^\dagger\}=H-q_i\geq0$, with central charges $q_i$. Then the BPS bound becomes $H=\max_i(q_i)$, and states that saturate this bound are called BPS states.}.
A subspace spanned by BPS states is called \emph{BPS subspace} and we denote it as $\mathcal{H}^{\rm BPS}_f$ within $\mathcal{H}_f$. In particular, the BPS subspace $\mathcal{H}^{\rm BPS}_f$ is a set of states annihilated by both $Q_{\rm SYK}$ and $Q_{\rm SYK}^\dagger$.

It was shown in \cite{Kanazawa:2017dpd} that the nilpotency of $Q$ and $Q^\dagger$ induces a further orthogonal decomposition,
\begin{equation}\label{eq:Hdecompose}
    \mathcal{H}_f=\mathcal{H}_f^+\oplus\mathcal{H}_f^-\oplus\mathcal{H}^{\rm BPS}_f\,,
\end{equation}
where $\mathcal{H}_f^+$ and $\mathcal{H}_f^-$ are non-BPS subspaces spanned by states $\Psi$ satisfying
\begin{equation}
Q_{\rm SYK}\Psi=0,\quad Q_{\rm SYK}^\dagger\Psi\neq0,    
\end{equation}
and
\begin{equation}
Q_{\rm SYK}\Psi\neq0,\quad Q_{\rm SYK}^\dagger\Psi=0,
\end{equation}
respectively. It was also shown in \cite{Kanazawa:2017dpd} that spectral statistics in each sector $\mathcal{H}_f^{\pm}$ are governed by random matrix theory, which means that the $\mathcal{N}=2$ SYK model exhibits quantum chaos in each non-BPS sector. 

As a concrete illustration of this quantum chaotic behavior of the $\mathcal{N}=2$ SYK model, let us take $N=12$ and consider $\mathcal{H}^+_{f=6}$, for example. This sector is described by the random matrix theory of the Gaussian unitary ensemble (GUE) \footnote{For the classification of general pairs of $(N,f)$, see (6.13) in \cite{Kanazawa:2017dpd}.}.

\begin{figure}[t]
\centering\includegraphics[width=3.37in]{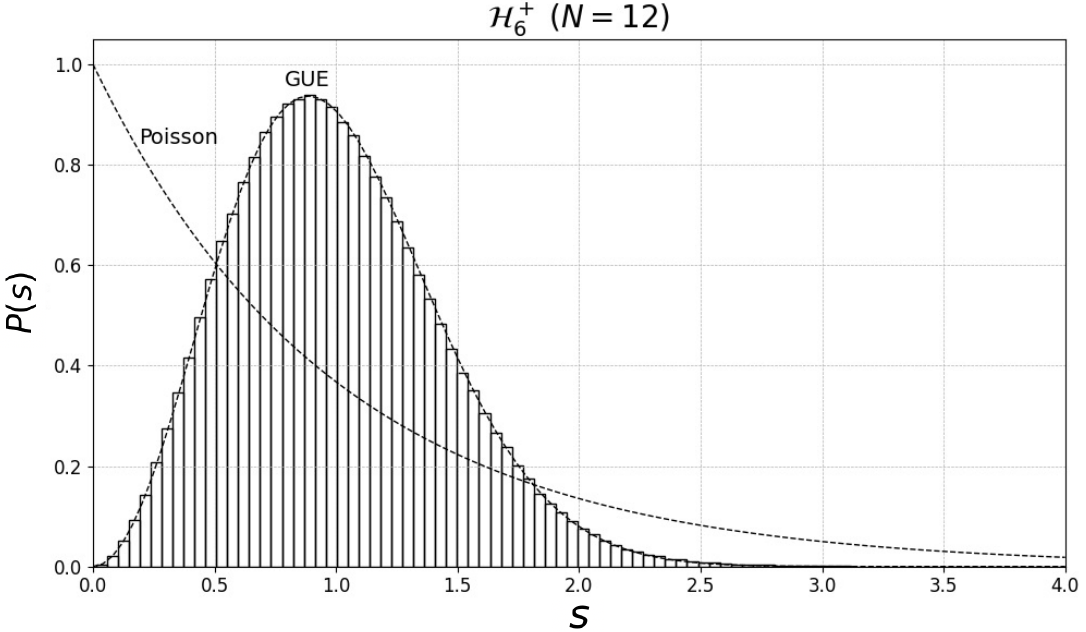}
	  \caption{The nearest level spacing distribution for $\mathcal{H}^+_{f=6}$ in $\mathcal{N}=2$ SYK model. The histogram is obtained by averaging over $4000$ realizations,
      while the dotted curves represent the Poisson and GUE Wigner-Dyson predictions. It agrees well with the GUE prediction.}\label{fig:Q3spacing}
\end{figure}

{
To characterize correlations between neighboring energy levels, we use two standard local spectral diagnostics: the averaged gap ratio $\langle r\rangle$ and the nearest-level-spacing distribution $P(s)$. Let $E_1\leq E_2\leq\cdots\leq E_D$
be the ordered eigenvalues of the Hamiltonian in the sector under consideration, and define the consecutive gaps $\delta_i=E_{i+1}-E_i$. The averaged gap ratio, which compares the relative sizes of two consecutive gaps, is
\begin{equation}\label{eq:rdef}
    \langle r\rangle
    =\overline{\frac{1}{D-2}\sum_{i=1}^{D-2}r_i}\,,\qquad
    r_i=\frac{\min(\delta_i,\delta_{i+1})}
    {\max(\delta_i,\delta_{i+1})}\,.
\end{equation}
The nearest-level-spacing distribution $P(s)$ is the probability density of the normalized spacing between adjacent eigenvalues. To obtain normalized spacings, we unfold the spectrum so that the local mean spacing is unity. Denoting the unfolded eigenvalues by $x_i$, $P(s)$ is defined as
\begin{equation}\label{eq:Psdef}
    s_i=x_{i+1}-x_i\,,\qquad
    P(s)=\overline{\frac{1}{D-1}\sum_{i=1}^{D-1}
    \delta(s-s_i)}\,.
\end{equation}
Here and below, the overline denotes the ensemble average. Generic integrable systems show a Poisson distribution $P(s)=e^{-s}$, whereas quantum chaotic systems show a Wigner-Dyson (WD) distribution, which is $P(s)=\frac{32}{\pi^2}s^2e^{-4s^2/\pi}$ for GUE. Correspondingly, $\langle r\rangle=2\log 2-1\approx0.386$ for Poisson and $\langle r\rangle\approx0.599$ for GUE \cite{Atas_2013}.
}
The result for the level-spacing distribution in $\mathcal{H}^+_{f=6}$ is shown in Fig. \ref{fig:Q3spacing}. It agrees well with the GUE WD distribution, indicating the quantum chaos in $\mathcal{H}^+_{f=6}$. We also find $\langle r\rangle = 0.598\cdots$, consistent with GUE.

\begin{figure}[h]
\centering\includegraphics[width=3.37in]{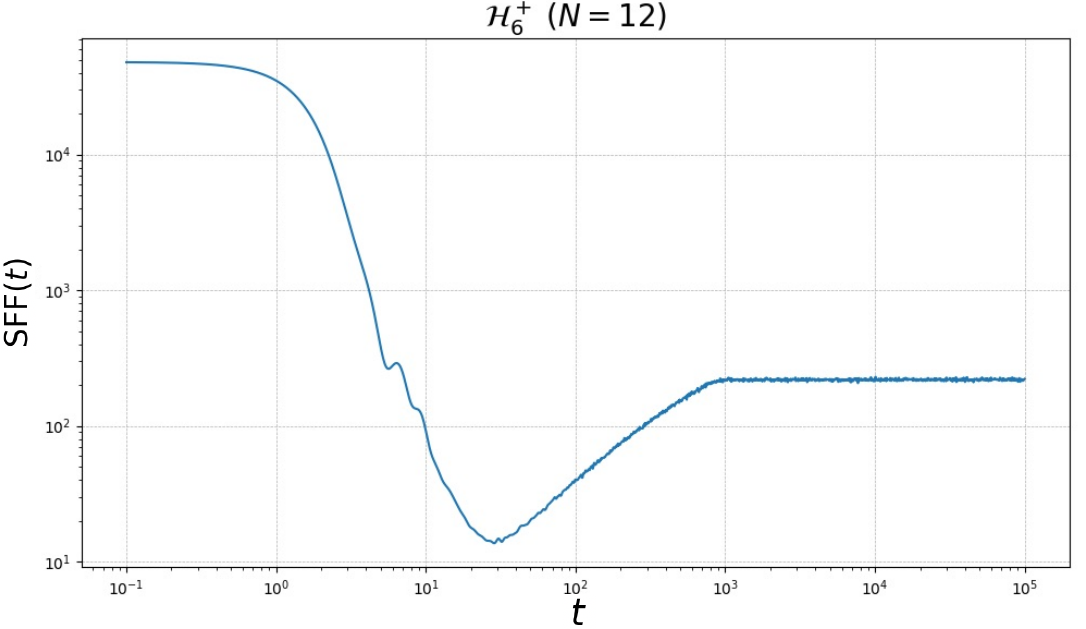}
	  \caption{The SFF for $\mathcal{H}^+_{f=6}$ in $\mathcal{N}=2$ SYK model averaged over $4000$ realizations. It exhibits a slope-ramp-plateau structure reflecting the chaotic spectrum.}\label{fig:Q3SFF}
\end{figure}

The spectral form factor (SFF) is another useful diagnostic of quantum chaos, as it probes long-range spectral correlations.
\begin{equation}\label{eq:Qcom}
    \text{SFF}(t)=\sum_{i,j}e^{-i(E_i-E_j)t}\,.
\end{equation}
Fig. \ref{fig:Q3SFF} shows a plot of the SFF computed using the spectrum in $\mathcal{H}^+_{f=6}$. This characteristic behavior referred to as the \emph{slope-ramp-plateau} is a sharp signal of quantum chaos. 

To investigate a possible chaos-integrability transition, we study the finite $N$ crossover induced by a deformation in terms of a supercharge
\begin{equation}
    Q_{\rm com}\equiv\sum_{a=1}^{N/3}C_{a}\psi_{3a-2}\psi_{3a-1}\psi_{3a}\,,
\end{equation}
with random coupling constants $C_a$ drawn from the same ensemble as $C_{ijk}$.
Here, the subscript ``com'' refers to the fact that the corresponding Hamiltonian decomposes into a sum of mutually commuting blocks, making this theory manifestly integrable. In fact, using the occupation number $n_i$ at the $i$-th fermion site, we can write the Hamiltonian $H_{\rm com}\equiv\{Q_{\rm com},Q_{\rm com}^{\dagger}\}$ as
\begin{equation}\label{eq:Hcomn}
\begin{split}
    H_{\rm com}=\sum_{a=1}^{N/3}|C_a|^2&\left[n_{3a-2}n_{3a-1}n_{3a}\right.\\
    &\left.+(1-n_{3a-2})(1-n_{3a-1})(1-n_{3a})\right]\,,
\end{split}
\end{equation}
demonstrating that the spectrum is given by a linear combination of $|C_{a}|^2$'s. This can be viewed as a supersymmetric generalization of \cite{Gao:2023gta}.

In the rest of the paper, we take $N$ to be a multiple of $3$ to appropriately define $Q_{\rm com}$, and we would like to consider a theory that has a supercharge
\begin{equation} \label{eq:Qg}
    Q_g\equiv Q_{\rm com} + 2^{-gN/2}Q_{\rm SYK}\,,
\end{equation}
with a coupling constant $g$ \footnote{The exponential $N$ dependence of the coefficient is chosen in analogy with the generalized Rosenzweig-Porter model \cite{RPmodel,Kravtsov_2015}, where the strength of off-diagonal mixing scales as a power of the Hilbert-space dimension. In our numerics, the smooth transition occurs around $g=1$ due to this $N$ dependence.}. Specifically, the Hamiltonian of the deformed model is defined by
\begin{equation}
    H_g\equiv\{Q_g,Q_g^\dagger\}.
\end{equation}
Therefore, the positive semi-definiteness of the Hamiltonian \eqref{eq:Hbound} continues to hold, and the decomposition of the Hilbert space \eqref{eq:Hdecompose} continues to hold for the deformed model \eqref{eq:Qg}, since the deformation by $Q_{\rm com}$ preserves the nilpotency and the fermion number \eqref{eq:f}. 

\begin{figure}[t]
\centering\includegraphics[width=3.37in]{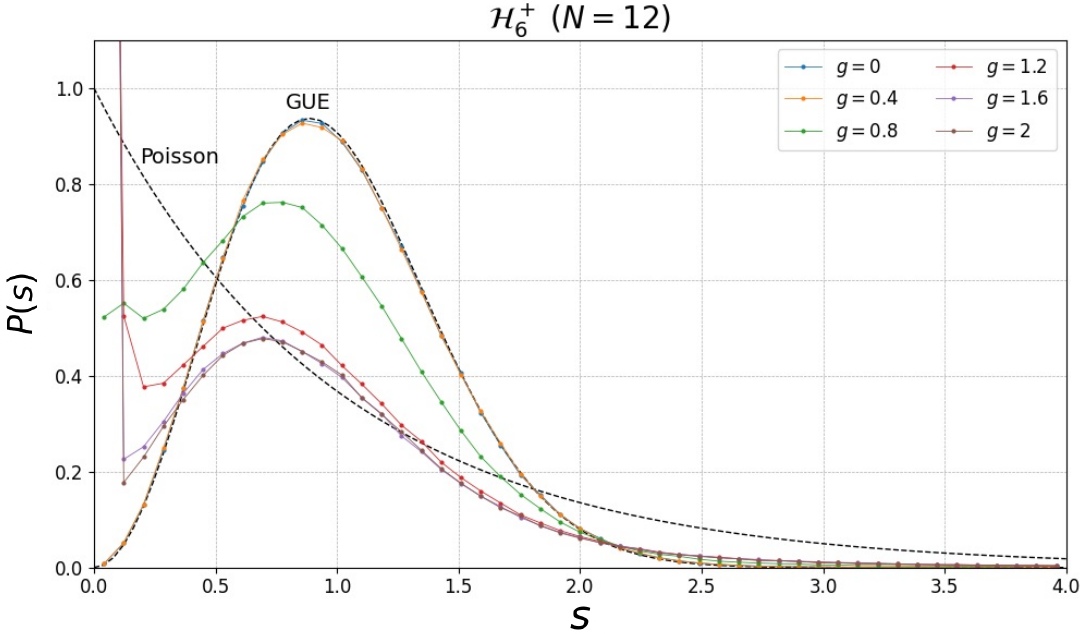}
	  \caption{The nearest level spacing distribution for the deformed model \eqref{eq:Qg}. Deviation of the distribution from GUE is shown as $g$ increases. The average is taken over $4000$ realizations for each value of $g$.}\label{fig:NonBPS_NLSD}
\end{figure}

\begin{figure}[t]
\centering\includegraphics[width=3.37in]{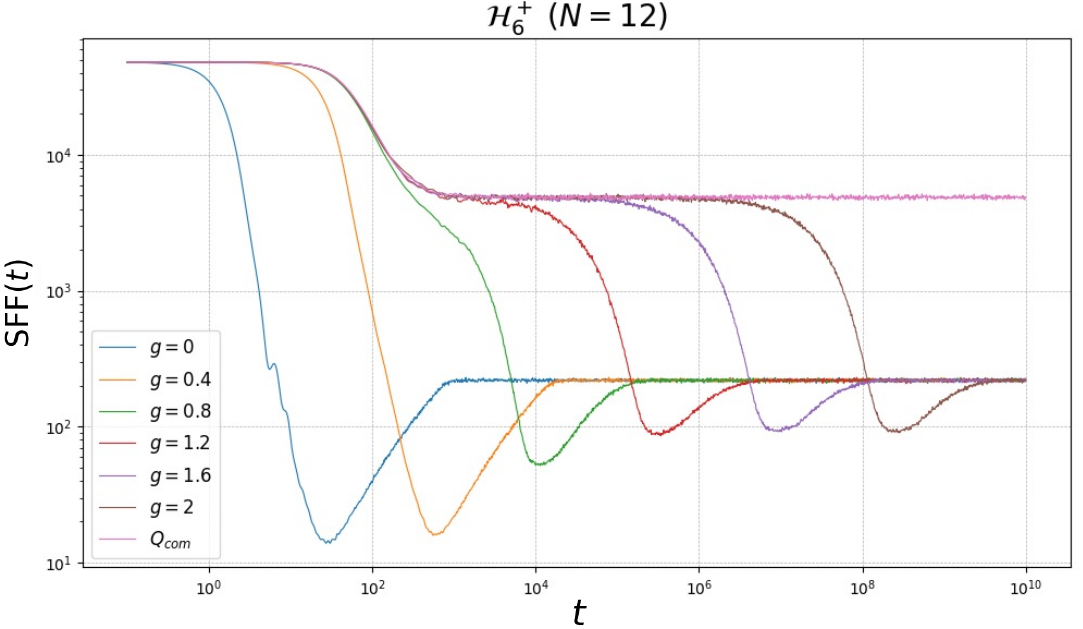}
	  \caption{The SFF for the deformed model \eqref{eq:Qg} averaged over $4000$ realizations for each $g$. The plateau for $Q_{\rm com}$ alone arises from its degeneracy in the spectrum. As $g$ increases, the ramp becomes small and approaches $Q_{\rm com}$, indicating a loss of chaotic behavior.}\label{fig:NonBPS_SFF}
\end{figure}

We can observe a chaos-integrability crossover in $\mathcal{H}^+_{f=6}$ as shown in Fig. \ref{fig:NonBPS_NLSD}. Since $H_{\rm com}$ is exactly solvable, its spectrum is highly structured and contains degeneracies. Consequently, the raw level-spacing distribution does not display a clean Poisson form \footnote{If one further resolves the Hilbert space using the additional conserved block quantum numbers of $H_{\rm com}$, the remaining level statistics are expected to become Poisson. However, these block quantum numbers are not preserved once the $Q_{\rm SYK}$ is turned on, and hence the same refinement cannot be employed for the deformed model \eqref{eq:Qg}.}. 
For the same reason, the SFF does not smoothly approach the integrable limit. Nevertheless, the disappearance of the ramp region in Fig. \ref{fig:NonBPS_SFF} indicates a loss of quantum chaos.

These observations motivate us to raise the question whether we may observe a similar crossover in the BPS subspace, in which the usual analysis based on spectral statistics does not simply apply. It turns out that we can see a clear crossover in the BPS subspace by making use of the notion of \emph{BPS chaos}.


\vspace{4pt}
\noindent \emph{3.~ \textbf{BPS chaos and fortuity}.—}
We would like to diagnose the chaos-integrability crossover by focusing solely on the BPS subspace of the deformed model \eqref{eq:Qg}. Since all BPS states are exactly degenerate zero-energy ground states, ordinary level statistics of the Hamiltonian cannot be used to probe chaos within this subspace. Nevertheless, the notion of \emph{BPS chaos}, originally motivated by LMRS \cite{Lin:2022rzw,Lin:2022zxd} and further developed in \cite{Chen:2024oqv}, provides a way to characterize chaotic behavior among BPS states.

An important question is how random-matrix behavior can arise among BPS states whose energies are exactly protected. A proposed explanation is the mechanism of \emph{chaos invasion} \cite{Chen:2024oqv}, which depends on the idea of \emph{fortuity} \cite{Chang:2024zqi}.

To formulate this idea, let us consider a sequence of theories with increasing $N$, in which the fermions and couplings already present at smaller $N$ are retained \cite{Chang:2024lxt}. Since the Hilbert spaces at different $N$ are different, continuing a BPS state does not mean keeping the same state vector. Rather, it means uplifting the corresponding $Q$-cohomology class \footnote{For $H=\{Q,Q^\dagger\}$, the Hodge decomposition implies that every nontrivial $Q$-cohomology class has a unique representative annihilated by both $Q$ and $Q^\dagger$. Hence the $Q$-cohomology is isomorphic to the BPS subspace. The same statement holds for $Q^\dagger$-cohomology.} to a nontrivial class in the theory at larger $N$. Some BPS states admit such an uplift repeatedly and therefore remain BPS at arbitrarily large $N$; these are called \emph{monotone} states. For other BPS states, every possible sequence of uplifts terminates at a finite value of $N$, beyond which the corresponding state is no longer BPS; these are called \emph{fortuitous} states. Thus, the distinction describes whether BPS protection persists as the theory is enlarged, rather than a property defined within a single fixed-$N$ theory.

Fortuitous states have been conjectured to describe typical microstates of supersymmetric black holes, whereas monotone states are associated with smooth horizonless geometries \cite{Chang:2024zqi}. This motivates the expectation that fortuitous BPS states should exhibit stronger signatures of quantum chaos than monotone ones. The \emph{chaos invasion} picture qualitatively describes this expectation: a fortuitous BPS state at a given $N$ originates from a finite-energy non-BPS state at larger $N$. Therefore, if the original finite-energy states are chaotic, their random matrix character may persist after they enter the BPS subspace.

Following the LMRS proposal \cite{Lin:2022rzw,Lin:2022zxd}, chaos within a BPS subspace, or \emph{BPS chaos}, can be probed by restricting a generic simple operator $O$ to the BPS subspace. In a basis $\{|a\rangle\}$ of BPS states, the resulting operator $O_{\rm BPS}$, defined in Eq.~\eqref{eq:OBPS}, has matrix elements $(O_{\rm BPS})_{ab}=\langle a|O|b\rangle$.
Random matrix statistics that arise in this restricted matrix are interpreted as a chaotic property of the BPS subspace. \footnote{Here, BPS chaos is diagnosed through the spectral
statistics of the restricted operator $O_{\rm BPS}$. This
diagnostic is distinct from operator growth under Hamiltonian
time evolution, which is not studied in this work.}

Motivated by this criterion, we will diagnose chaos in the BPS subspace of the deformed SYK model by studying the spectral statistics of a projected simple operator.

\vspace{4pt}
\noindent \emph{4.~ \textbf{Chaos-integrability transition in the BPS subspace}.—}
A key observation here is that the BPS states of both $Q_{\rm SYK}$ and the integrable deformation $Q_{\rm com}$ are all fortuitous. 
The chaos invasion picture suggests that the spectral statistics of finite-energy states can be inherited by fortuitous BPS states. Therefore, $Q_{\rm SYK}$ is expected to transmit chaotic random-matrix behavior to the BPS subspace, whereas the deformation $Q_{\rm com}$ may instead transmit Poisson-like statistics from its integrability.
This motivates us to investigate whether the deformed model \eqref{eq:Qg} exhibits a chaos-integrability transition directly within its BPS subspace.

The fortuitous nature of these BPS states can be understood as follows. 
For $Q_{\rm SYK}$, the BPS subspace exhibits the $R$-charge concentration discussed in \cite{Chang:2024lxt}: BPS states are concentrated within an $\mathcal{O}(1)$ range around $\mathcal{H}_{f=N/2}$. An uplift of $Q$-cohomology preserves the fermion number, hence cannot remain within the concentration window around $f=N/2$ for arbitrarily large $N$. Therefore, these BPS states are fortuitous.
For $Q_{\rm com}$, we can see from \eqref{eq:Hcomn}, that a state is BPS only when each block (labeled by $a$) contains either one or two fermions. Therefore, the BPS subspace is supported in the range $\frac{N}{3}\leq f\leq \frac{2N}{3}$ for fixed $N$. Although this distribution is not concentrated, its states likewise admit no infinite BPS uplift. A $Q$-cohomology uplift adds an empty block giving a positive contribution in \eqref{eq:Hcomn}. Hence this uplift does not remain BPS, and these states are also fortuitous.

For simplicity, we focus on $N=12$ and the sector $\mathcal{H}^{\rm BPS}_{f=6}$, in which we find that the dimension of the BPS subspace remains unchanged throughout the interpolation in $g$.
To see a clear transition, we need to choose a probe operator $O$ that does not have a degenerate spectrum in $\mathcal{H}^{\rm BPS}_{f=6}$. A natural choice for such an operator is
\begin{equation}\label{eq:O}
O=\sum_{i=1}^{N}A_{i}\psi_{i}^{\dagger}\psi_{i}\,,
\end{equation}
where $A_{i}$ are Gaussian random real variables with zero mean and unit variance. \footnote{The randomness of $A_i$ alone does not generate random-matrix statistics, since the same probe ensemble yields Poisson-like statistics at the integrable endpoint. A fixed probe such as $O=\psi_{1}^{\dagger}\psi_{1}$ gives the same qualitative crossover, although its degeneracies make the spectral statistics less clean.}

Now, defining $O_{\text{BPS},\,f=6}$ as the operator $O$ projected onto $\mathcal{H}^{\rm BPS}_{f=6}$, we are ready to investigate the chaos-integrability crossover in $\mathcal{H}^{\rm BPS}_{f=6}$. Using the spectrum of $O_{\rm BPS,\,f=6}$, we can compute $\langle r\rangle$, the nearest level spacing distribution, and the SFF, in the same way as was done in Sec. 2 with the spectrum of the Hamiltonian.

\begin{figure}[t]
\centering\includegraphics[width=\columnwidth]{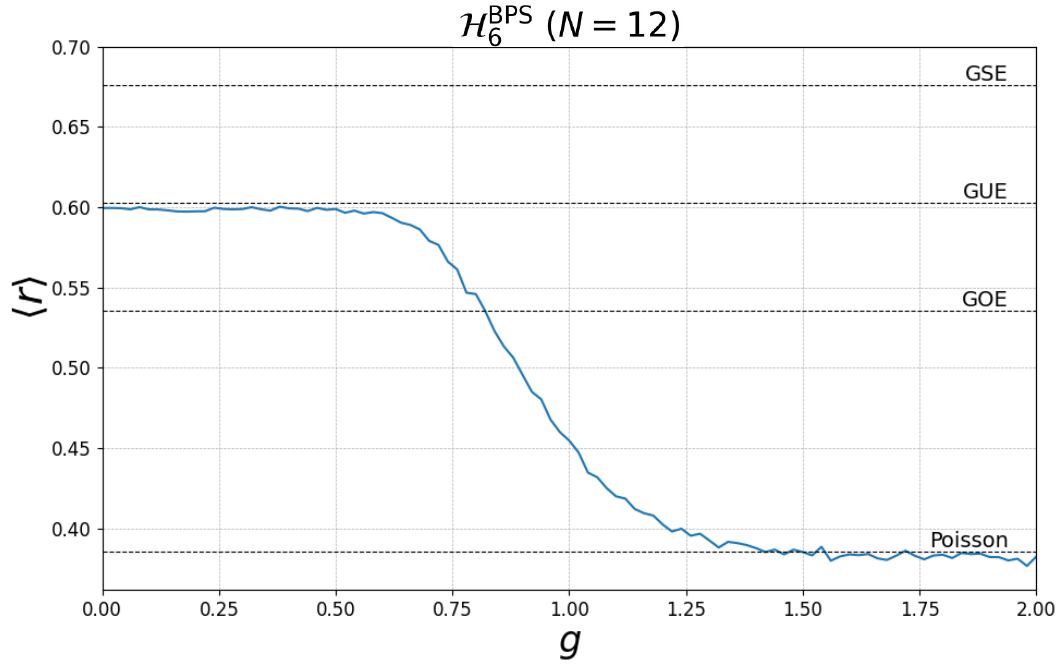}
      \caption{Dependence of the averaged gap ratio $\langle r\rangle$ defined in Eq.~\eqref{eq:rBPS} on the deformation parameter $g$, computed from the spectrum of $O_{\rm BPS,\,f=6}$ for $N=12$. The average is taken over $500$ realizations. The result exhibits a smooth crossover from the GUE value to the Poisson value around $g=1$.
}\label{fig:BPS_r}
\end{figure}

\begin{figure}[t]
\centering\includegraphics[width=\columnwidth]{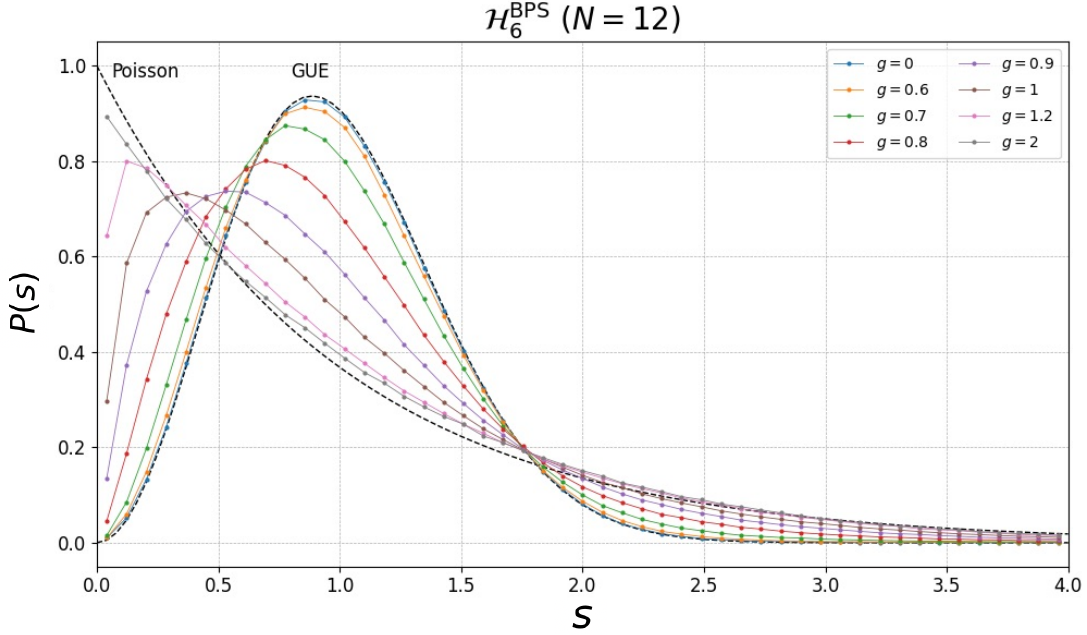}
	  \caption{Nearest level-spacing distribution $P(s)$ defined in Eq.~\eqref{eq:PBPS} for the spectrum of the projected operator $O_{\rm BPS,\,f=6}$ in the deformed model \eqref{eq:Qg} for $4000$ realizations. The distribution shows a crossover from GUE to Poisson around $g=1$.}\label{fig:BPS_NLSD}
\end{figure}

{
More explicitly, let $\lambda_1\leq\lambda_2\leq\cdots\leq\lambda_D$ be the ordered eigenvalues of $O_{\rm BPS,\,f=6}$, and define $\delta_i^{(O)}=\lambda_{i+1}-\lambda_i$.
We can define the gap ratio for this spectrum as
\begin{equation}\label{eq:rBPS}
    \langle r\rangle
    =\overline{\frac{1}{D-2}\sum_{i=1}^{D-2}r_i^{(O)}}\,,\qquad
    r_i^{(O)}=\frac{\min(\delta_i^{(O)},\delta_{i+1}^{(O)})}
    {\max(\delta_i^{(O)},\delta_{i+1}^{(O)})}\,.
\end{equation}
Denoting the unfolded eigenvalues of $O_{\rm BPS,\,f=6}$ by
$x_i^{(O)}$, we can also define the nearest level spacing distribution 
\begin{equation}\label{eq:PBPS}
    s_i^{(O)}=x_{i+1}^{(O)}-x_i^{(O)}\,,\qquad
    P(s)=\overline{\frac{1}{D-1}\sum_{i=1}^{D-1}
    \delta(s-s_i^{(O)})}\,.
\end{equation}
Figures~\ref{fig:BPS_r} and \ref{fig:BPS_NLSD} both show a clear crossover from GUE to Poisson statistics around $g=1$.
}


\begin{figure}[t]
\centering\includegraphics[width=\columnwidth]{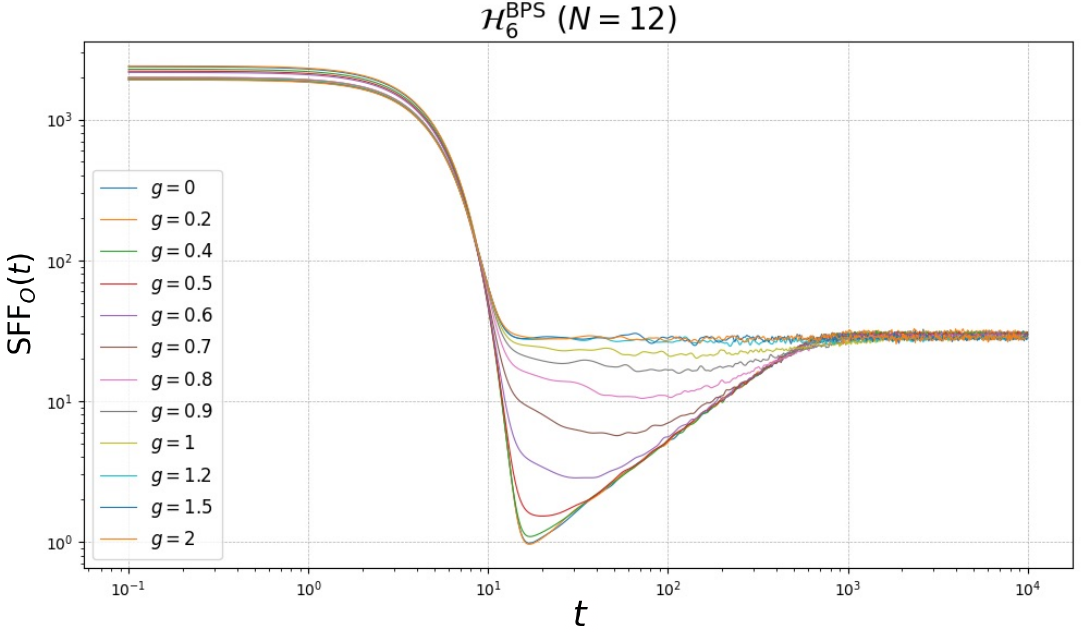}
	  \caption{The filtered spectral form factor $\operatorname{SFF}_{O}(t)$ defined in Eq.~\eqref{eq:SFFO}, averaged over $4000$ realizations, with $\sigma=0.2$. The ramp region seems to disappear around $g=1.2$, suggesting the chaos-integrability transition.}\label{fig:BPS_SFF}
\end{figure}

{
Fig.~\ref{fig:BPS_SFF} shows the SFF defined directly from the eigenvalues of $O_{\rm BPS,\,f=6}$ by
\begin{align}\label{eq:SFFO}
    \operatorname{SFF}_{O}(t)
    &=\overline{\left|\sum_{i}
    e^{-\lambda_i^2/(2\sigma^2)}
    e^{-i\lambda_i t}\right|^2}\,.
\end{align}
Here, the subscript $O$ represents the dependence on our choice of the operator.
The Gaussian filter \cite{Gharibyan:2018jrp} is introduced to suppress edge effects and focus on bulk spectral correlations.
Note that $t$ is conjugate to differences between the eigenvalues of $O_{\rm BPS,\,f=6}$ and does not represent physical time evolution generated by $H_g$. Therefore, this SFF cannot, in general, be identified with a dynamical correlation function of $O$ and instead probes spectral correlations of $O_{\rm BPS,\,f=6}$.

In Fig.~\ref{fig:BPS_SFF}, we observe a disappearance of the ramp region, suggesting a chaos-integrability transition.
}

\vspace{4pt}
\noindent \emph{5.~\textbf{Conclusion and Outlook}.—}
In this paper, we demonstrated that a chaos-integrability crossover can be observed directly within the BPS subspace of a deformed $\mathcal N=2$ SYK model, using the diagnostic framework of BPS chaos. In particular, the projected operator spectrum transitions from GUE to Poisson statistics as the deformation parameter is varied. 
Since both endpoints contain only fortuitous BPS states, our results indicate that fortuitous BPS states can inherit qualitatively different spectral characteristics depending on the structure of the finite-energy spectrum from which they descend.

The holographic interpretation of this crossover is an important open question. Since the $\mathcal N=2$ SYK model is described at low energies by the super-Schwarzian or JT supergravity \cite{Turiaci:2023jfa}, the disappearance of random-matrix behavior suggests a breakdown of the black-hole-like chaotic description as the model approaches the integrable regime. More broadly, in supersymmetric holographic systems where a bulk description can be followed along a deformation, BPS chaos diagnostics may provide boundary probes of changes in the dominant semiclassical saddle, for example between black-hole-like and horizonless descriptions.
A relation between wormhole length and an appropriately defined Krylov complexity in the BPS subspace of the $\mathcal{N}=2$ double-scaled SYK was discussed in \cite{Aguilar-Gutierrez:2025sqh} and may provide useful insight in this regard.

Whether this finite $N$ crossover sharpens into a genuine transition in the large $N$ limit remains an open question. It would be worthwhile to investigate whether an analytic large $N$ description can be developed for our model, perhaps in line with the analysis of the deformed SYK model in \cite{Garcia-Garcia:2017bkg}. Another promising direction is to study the double-scaling limit \cite{Berkooz:2020xne} and formulate the model in terms of a chord path integral established in the analysis of chaos-integrability transition in ordinary double-scaled SYK model \cite{Berkooz:2024evs, Berkooz:2024ofm, Aguilar-Gutierrez:2026jjv}.

Although we have used the spectral form factor of a projected operator as a diagnostic of BPS chaos, it would be interesting to ask whether other probes can be adapted to detect the same crossover within the BPS sector. Possible candidates include OTOC \cite{Sekino:2008he,Maldacena:2015waa}, entanglement entropy \cite{Wang:2003zdo}, relaxation fluctuations of correlation functions \cite{Pathak:2024kki}, Krylov complexity \cite{Balasubramanian:2022tpr, Bhattacharjee:2022ave}, and (generating function of) spectral complexity \cite{Iliesiu:2021ari,Miyaji:2025yvm,Miyaji:2025jxy}. For extensions to more general supersymmetric models, the spectrum of the Berry curvature may provide an intrinsic diagnostic of BPS chaos \cite{Chen:2026vml}. 

Finally, clarifying the precise relation between fortuity and chaos is another important problem. The present model provides a useful example in this regard: $Q_{\rm com}$ has only fortuitous BPS states despite being integrable. It would be interesting to investigate whether the sharpness of the $R$-charge distribution of BPS states is correlated with quantum chaos, since for $Q_{\rm SYK}$, the BPS states are concentrated within $\mathcal O(1)$ charge window, whereas for $Q_{\rm com}$ they are supported over $\mathcal O(N)$ range, $N/3\leq f\leq 2N/3$. Another model with fortuitous BPS states distributed over $\mathcal O(N)$ range of $R$-charge, while being integrable, was proposed in \cite{Chen:2025sum}.

\vspace{8pt}
\begin{acknowledgments}
\noindent \emph{Acknowledgments:}
We are grateful to Marcel R. R. Hughes for valuable comments and discussions, and also thank Yichao Fu, Shan-Ming Ruan, and Masaki Shigemori for insightful discussions.
LM and SS are supported by JST SPRING, grant number JPMJSP2125, ``THERS Make New Standards Program for the Next Generation Researchers".
\end{acknowledgments}

\appendix

\section*{Appendix A. Integrability of $Q_{\rm com}$}
\begin{figure}[h]
\centering\includegraphics[width=3.37in]{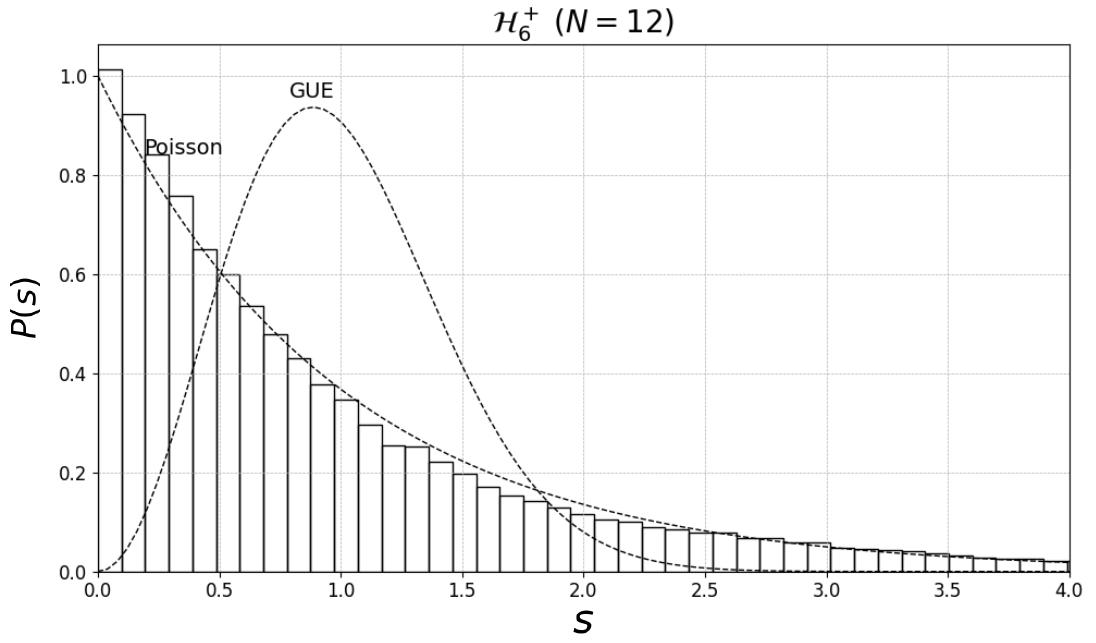}
	  \caption{The nearest level spacing distribution for $\mathcal{H}^+_{f=6}$ in $H_{\rm com}$ alone with degeneracy removed. The histogram is obtained by averaging over $10000$ realizations. It agrees well with the Poisson distribution, indicating its integrability.}\label{fig:level_spacing_Qcom}
\end{figure}

In the main text, the nearest level spacing distribution for $H_g$ did not show clear Poisson distribution at large $g$, as shown in Fig. \ref{fig:NonBPS_NLSD}. This is due to the degeneracy in the spectrum of $H_{\rm com}$. For completeness, we show in Fig. \ref{fig:level_spacing_Qcom} that the nearest level spacing distribution for $H_{\rm com}$ alone indeed exhibits Poisson-like statistics, if we remove the degeneracy.

\bibliography{references.bib}

\end{document}